\documentclass[british,english,twocolumn,english,aip,reprint]{revtex4-1}
\usepackage{amsmath}
\usepackage{newtxmath}
\usepackage[LGR,T1]{fontenc}
\usepackage[latin9]{inputenc}
\usepackage{array}
\usepackage{booktabs}
\usepackage{textcomp}
\usepackage{graphicx}

\makeatletter

\DeclareRobustCommand{\greektext}{%
  \fontencoding{LGR}\selectfont\def\encodingdefault{LGR}}
\DeclareRobustCommand{\textgreek}[1]{\leavevmode{\greektext #1}}
\ProvideTextCommand{\~}{LGR}[1]{\char126#1}

\providecommand{\tabularnewline}{\\}

\newcommand*{\citen}[1]{%
  \begingroup
    \romannumeral-`\x 
    \setcitestyle{numbers}%
    \cite{#1}%
  \endgroup   
}

\makeatother

\usepackage{babel}
\begin{document}
\title{Plasma-assisted molecular beam epitaxy of NiO on GaN(00.1)}
\author{Melanie Budde}
\email{budde@pdi-berlin.de}

\affiliation{Paul-Drude-Institut f\"ur Festk\"orperelektronik, Leibniz-Institut im
Forschungsverbund Berlin e.V., Hausvogteiplatz 5-7, 10117 Berlin,
Germany}
\author{Thilo Remmele}
\affiliation{Leibniz-Institut f\"ur Kristallz\"uchtung, Max-Born-Str. 2, 12489 Berlin,
Germany}
\author{Carsten Tschammer}
\affiliation{Paul-Drude-Institut f\"ur Festk\"orperelektronik, Leibniz-Institut im
Forschungsverbund Berlin e.V., Hausvogteiplatz 5-7, 10117 Berlin,
Germany}
\author{Johannes Feldl}
\affiliation{Paul-Drude-Institut f\"ur Festk\"orperelektronik, Leibniz-Institut im
Forschungsverbund Berlin e.V., Hausvogteiplatz 5-7, 10117 Berlin,
Germany}
\author{Philipp Franz}
\affiliation{Paul-Drude-Institut f\"ur Festk\"orperelektronik, Leibniz-Institut im
Forschungsverbund Berlin e.V., Hausvogteiplatz 5-7, 10117 Berlin,
Germany}
\author{Jonas L\"ahnemann}
\affiliation{Paul-Drude-Institut f\"ur Festk\"orperelektronik, Leibniz-Institut im
Forschungsverbund Berlin e.V., Hausvogteiplatz 5-7, 10117 Berlin,
Germany}
\author{Zongzhe Cheng}
\affiliation{Paul-Drude-Institut f\"ur Festk\"orperelektronik, Leibniz-Institut im
Forschungsverbund Berlin e.V., Hausvogteiplatz 5-7, 10117 Berlin,
Germany}
\author{Michael Hanke}
\affiliation{Paul-Drude-Institut f\"ur Festk\"orperelektronik, Leibniz-Institut im
Forschungsverbund Berlin e.V., Hausvogteiplatz 5-7, 10117 Berlin,
Germany}
\author{Manfred Ramsteiner}
\affiliation{Paul-Drude-Institut f\"ur Festk\"orperelektronik, Leibniz-Institut im
Forschungsverbund Berlin e.V., Hausvogteiplatz 5-7, 10117 Berlin,
Germany}
\author{Martin Albrecht}
\affiliation{Leibniz-Institut f\"ur Kristallz\"uchtung, Max-Born-Str. 2, 12489 Berlin,
Germany}
\author{and Oliver Bierwagen}
\email{bierwagen@pdi-berlin.de}

\affiliation{Paul-Drude-Institut f\"ur Festk\"orperelektronik, Leibniz-Institut im
Forschungsverbund Berlin e.V., Hausvogteiplatz 5-7, 10117 Berlin,
Germany}
\date{\today\\
}
\begin{abstract}
\noindent The growth of NiO on GaN(00.1) substrates by plasma-assisted
molecular beam epitaxy under oxygen rich conditions was investigated
at growth temperatures between100\,$^{\circ}$C and 850\,$^{\circ}$C. Epitaxial growth
of NiO(111) with two rotational domains, with epitaxial relation {\normalsize{}$\mathrm{\mathrm{\mathrm{NiO}(1\bar{\mathrm{{1}}}0)}\:||\:\mathrm{\mathrm{GaN}(11.0)}}$}
and $\mathrm{\mathrm{\mathrm{NiO}\mathrm{(10\bar{\mathrm{{1}}})}\:||\:\mathrm{GaN(11.0)}}}$,
was observed by X\nobreakdash-ray diffraction (XRD) and confirmed
by in-situ reflection high-energy electron diffraction as well as
transmission electron microscopy (TEM) and electron backscatter diffraction.
With respect to the high lattice mismatch of 8.1\,\% and a measured
low residual tensile layer strain, growth by lattice matching epitaxy
or domain matching epitaxy is discussed. The morphology measured by
atomic force microscopy showed a grainy surface, probably arising
from the growth by the columnar rotational domains visible in TEM
micrographs. The domain sizes measured by AFM and TEM increase with
the growth temperature, indicating an increasing surface diffusion
length. Growth at 850\,$^{\circ}$C, however, involved local decomposition
of the GaN substrate that lead to an interfacial $\mathrm{\beta}$-Ga$\mathrm{_{2}}$O$\mathrm{_{3}}$($\bar{\mathrm{{2}}}$01)
layer and a high NiO surface roughness. Raman mesurements of the quasi-forbidden
one-phonon peak indicate increasing layer quality (decreasing defect
density) with increasing growth temperature.
\end{abstract}
\maketitle

\section{Introduction}

The transparent wide band gap (3.7 eV)\citep{Ohta} material nickel
oxide (NiO) crystallizes in the rock~salt crystal structure.\citep{Rao_NiO-prop}
Stoichiometric NiO is considered insulating, whereas unintentional
\textit{p}\nobreakdash-type conductivity is induced by Ni vacancies
formed under certain growth conditions.\citep{Rao_NiO-prop} Intentional
Li-doping has been shown to create \textit{p}\nobreakdash-type conductivity
in a controlled way.\citep{Rao_NiO-prop,Zhang2018} NiO is antiferromagnetic
with a N\'{e}el temperature of about 525~K,\citep{Schuler} below which
the crystal structure is slightly distorted with angles of > 90.1$^{\circ}$
instead of 90$^{\circ}$.\citep{Massidda_Distortion} However, this minor deviation
from the perfect cubic structure can be neglected for our study.

\noindent Its properties make NiO an interesting material for many
GaN-based applications: Heterojunction p-NiO/n-GaN diodes \citep{Li_Diode,Wang_Diode}
as well as normally-off operating heterojunction field-effect transistors
(HFETs)\citep{Suzuki_HFET} have been investigated, based on the equivalent
work function of NiO and GaN.\citep{Suzuki_HFET} Furthermore, NiO
has been used as a stable hydrogen reduction catalyst to enhance the
efficiency of GaN-based water splitting\citep{Kim_waterSplitting}
and protect the GaN from decomposition during this process. Finally,
NiO is a potential antiferromagnetic pinning layer in (GaN-based)
spintronic devices.\citep{Becker2017}.

\noindent So far, NiO on GaN has been grown by electron-beam evaporation,\citep{Reddy_interlayer}
RF magnetron sputtering,\citep{Wang_Diode} or metal organic chemical
vapor deposition (MOCVD).\citep{LoNigro_NiO_MOCVD} In addition, growth
by MOCVD on AlGaN is reported, reducing the lattice mismatch from
about 8~\% to 5~\%.\citep{Nigro_MOCVD_AlGaN,NiO_AlGaN} While plasma-assisted
molecular beam epitaxy (PA\nobreakdash-MBE) is a well-suited method
to grow high-quality oxide layers and interfaces, this method has
not been reported for the growth of NiO on GaN. We have previously
reported on the PA-MBE of NiO on MgO substrates for a wide range of
growth parameters (temperature, oxygen flux).\citep{MgO_NiO}\\
In the present study we investigate the properties of NiO thin films
grown by PA-MBE on GaN(00.1) substrates as a function of growth temperature.
In particular, the epitaxial relationship between the two materials,
the accomodation of the lattice mismatch and the effect of the growth
temperature on layer and interface quality are analyzed. Different
methods were used to measure the structural properties of the films,
allowing one to compare their advantages and disadvantages.

\section{Experiment}

\noindent NiO was grown by PA\nobreakdash-MBE on 2-inch (00.1)\nobreakdash-oriented
gallium nitride templates manufactured by ``Kyma Technologies''.
The templates consist of a sapphire substrate~($\alpha$-Al$_{2}$O$_{3}$),
an AlN buffer and a GaN layer. Substrate heating was improved by sputter-coated
Ti on the rough backside to absorb the radiation from the heating
filament. Five substrate heater temperatures, measured by a thermocouple
between substrate and heating filament (100\,$^{\circ}$C, 300\,$^{\circ}$C, 500\,$^{\circ}$C, 700\,$^{\circ}$C
and 850\,$^{\circ}$C) were used to investigate their influence on
the NiO layer. Throughout this report the corresponding samples are
named S100, S300, S500, S700, and S850, respectively. Ni was sublimed
from an effusion cell at a temperature of 1380~$^{\circ}$C, since staying
below the melting point of Ni (1455\,$^{\circ}$C\citep{Book_of_Chem+Phy})
is important to protect the Al$\mathrm{_{2}}$O$\mathrm{_{3}}$ crucible
of the effusion cell from breaking upon re-solidification of Ni.\citep{Thesis_Mares}
The resulting beam equivalent pressure (BEP), which is proportional
to the particle flux, was measured by a nude filament ion gauge positioned
at the substrate location and then removed before growth. For 1380~$^{\circ}$C,
a Ni BEP of about $\mathrm{8.5\cdot10^{-9}}$~mbar was measured for
all growth runs. Activated oxygen was provided by passing 0.5~standard
cubic centimeters per minute (sccm) of O$\mathrm{_{2}}$ through an
RF-plasma source run at an RF-power of 300~W, resulting in an oxygen
BEP of $\mathrm{1.5\cdot10^{-6}}$~mbar and oxygen-rich growth conditions.
In our growth study on MgO we found plasma activated oxygen to yield
significantly smoother NiO layers than molecular oxygen.\citep{MgO_NiO}
For growth, the Ni shutter was opened first and the oxygen was introduced
after 60s. The total growth time was 5000\,s for all samples. As
a comparison one sample without Ni preflow, named S500{*}, was grown
at 500\,$^{\circ}$C. During growth, the crystallinity and morphology was assessed
in-situ by reflection high-energy electron diffraction (RHEED).

\noindent After growth, different ex-situ methods were used to investigate
the layer properties. X\nobreakdash-ray diffraction (XRD) measurements
in a 4-circle lab-diffractometer using Cu\,K$\mathrm{\mathrm{\alpha}}$
radiation were conducted to investigate the epitaxial relationship
between substrate and layer by a combination of symmetric on axis
$\mathrm{\mathrm{\mathrm{2\theta-\omega}}}$ scans (out\nobreakdash-of\nobreakdash-plane
orientation) and a skew symmetric off axis $\mathrm{\mathrm{\varPhi}}$\nobreakdash-scan
(in\nobreakdash-plane relationship). The Laue oscillations visible
in the symmetric on axis $\mathrm{\mathrm{\mathrm{2\theta-\omega}}}$
scan around the NiO peak are used to determine the layer thickness.
In addition, synchrotron-based XRD $\mathrm{\mathrm{2\theta-\omega}}$
scans at grazing-incidence, measured at the PHARAO facility\citep{Jenichen2003}
as described in Ref.~{[}\citen{Cheng2017}{]}, were used to extract
the in-plane lattice parameter from the peak position of the NiO(2$\overline{\mathrm{2}}$0)
reflection referenced to that of the GaN(11.0) reflection.

\noindent The film domain structure and orientation was investigated
by electron backscatter diffraction (EBSD) measurements in a scanning
electron microscope fitted with an EDAX-TSL EBSD detector using an
acceleration voltage of 15 kV. Atomic force microscopy (AFM) in the
peak force tapping mode using a Bruker \textquotedblleft Dimension
edge\textquotedblright{} with the \textquotedblleft ScanAsyst\textquotedblright{}
technology was used to analyse the surface morphology and domain size.
Cross-sectional transmission electron microscopy (TEM) was used for
detailed investigations of the domain structure of the NiO layers,
the interface to the GaN substrate, and the strain state of the layers.
For this purpose, cross-section samples were prepared by polishing
on diamond lapping films, followed by Ar ion milling with 4 kV down
to 200 V acceleration voltage to minimize amorphous surface layers.
Finally, quality investigations were performed using the forbidden
one\nobreakdash-phonon\nobreakdash-Raman peak as we have previously
described for NiO on MgO.\citep{MgO_NiO} For this purpose, Raman
spectra were recorded at room temperature in backscattering geometry
using the 325~nm line of a Cd-He ion laser.

\section{Results and discussion}

\subsection{Epitaxial relation\label{subsec:Epitaxial-relationship}}

Fig.~\ref{fig:2Theta Combination}~(a)) shows the on-axis $\mathrm{2\theta-\omega}$
scan of sample S500 (grown at 500$^{\circ}$C) for $\mathrm{2\theta}$ between
15$^{\circ}$ and 100$^{\circ}$C. The presence of a strong NiO(111) peak and the absence
of other NiO peaks for this and all other investigated samples (shown
in Fig.~\ref{fig:2Theta Combination}~(b)) indicate the NiO out\nobreakdash-of\nobreakdash-plane
direction to be {[}111{]} and the corresponding out-of-plane epitaxial
relation:

\noindent 
\begin{equation}
\begin{array}{ccc}
 & \mathrm{NiO\left[111\right]\:||\:GaN\left[00.1\right]}\end{array}\label{eq:Grain1-1}
\end{equation}
The other (GaN, AlN and Al$\mathrm{_{2}}$O$\mathrm{_{3}}$) peaks
can be assigned to the template structure of the substrate.

\subsubsection*{
\begin{figure}[th]
\protect\centering{}\protect\includegraphics[width=8.5cm]{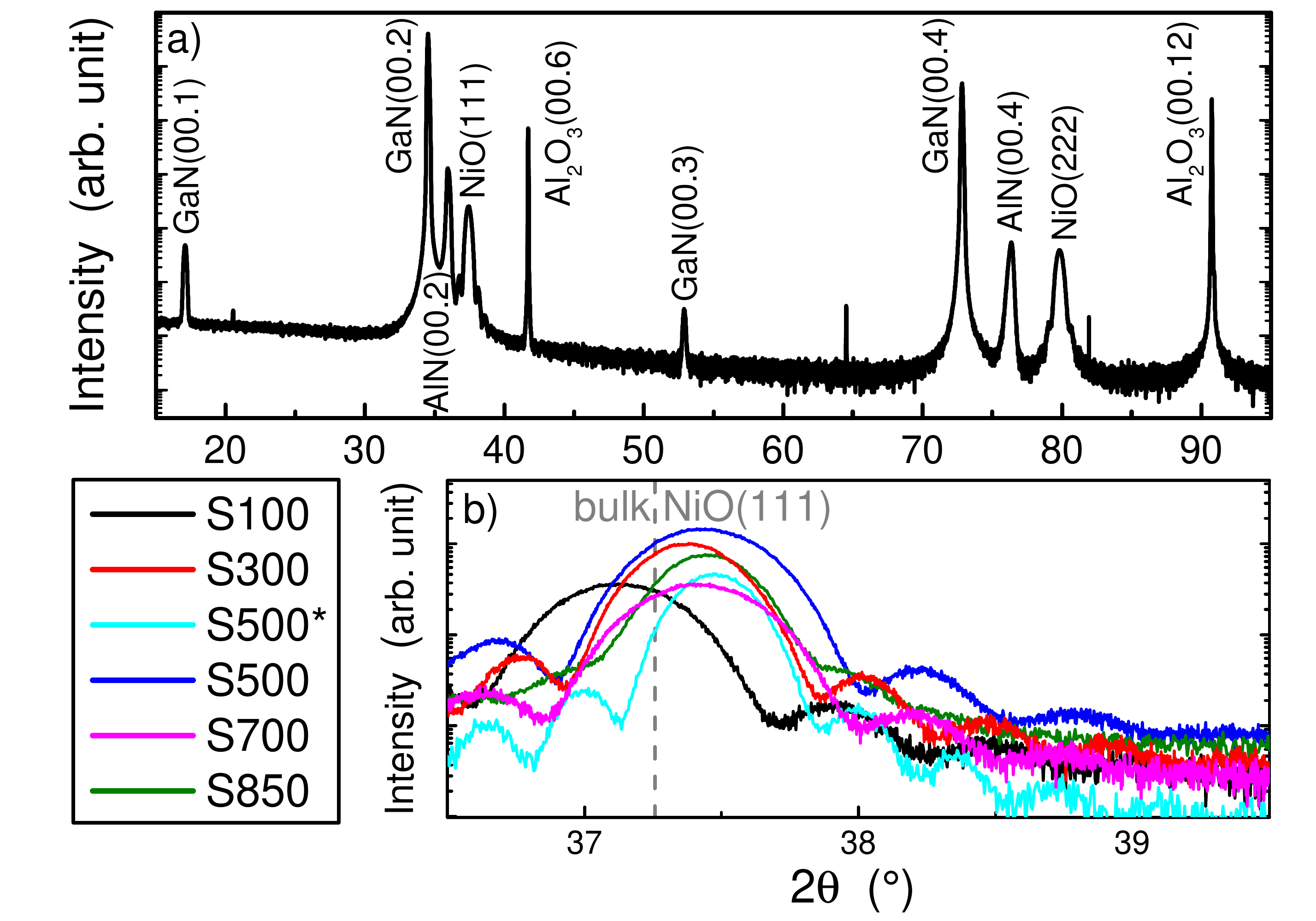}\protect\caption{a) Symmetric on-axis XRD $\mathrm{2\theta-\omega}$ scan of S500 showing
the NiO(111) and NiO(222) peak, as well as the substrate peaks (GaN,
AlN and Al$\mathrm{_{2}}$O$\mathrm{_{3}}$ due to the template structure
of the used substrate). b) Corresponding $\mathrm{2\theta-\omega}$
scans for all samples in the $\mathrm{2\theta}$ range from 36.5$^{\circ}$
to 39.5$^{\circ}$. (The NiO bulk position is calculated from the NiO(200) peak
measured on a bulk NiO foil.)\citep{MgO_NiO} \label{fig:2Theta Combination}}
\protect
\end{figure}
}

\begin{figure}[th]
\noindent \begin{centering}
\includegraphics[width=8cm]{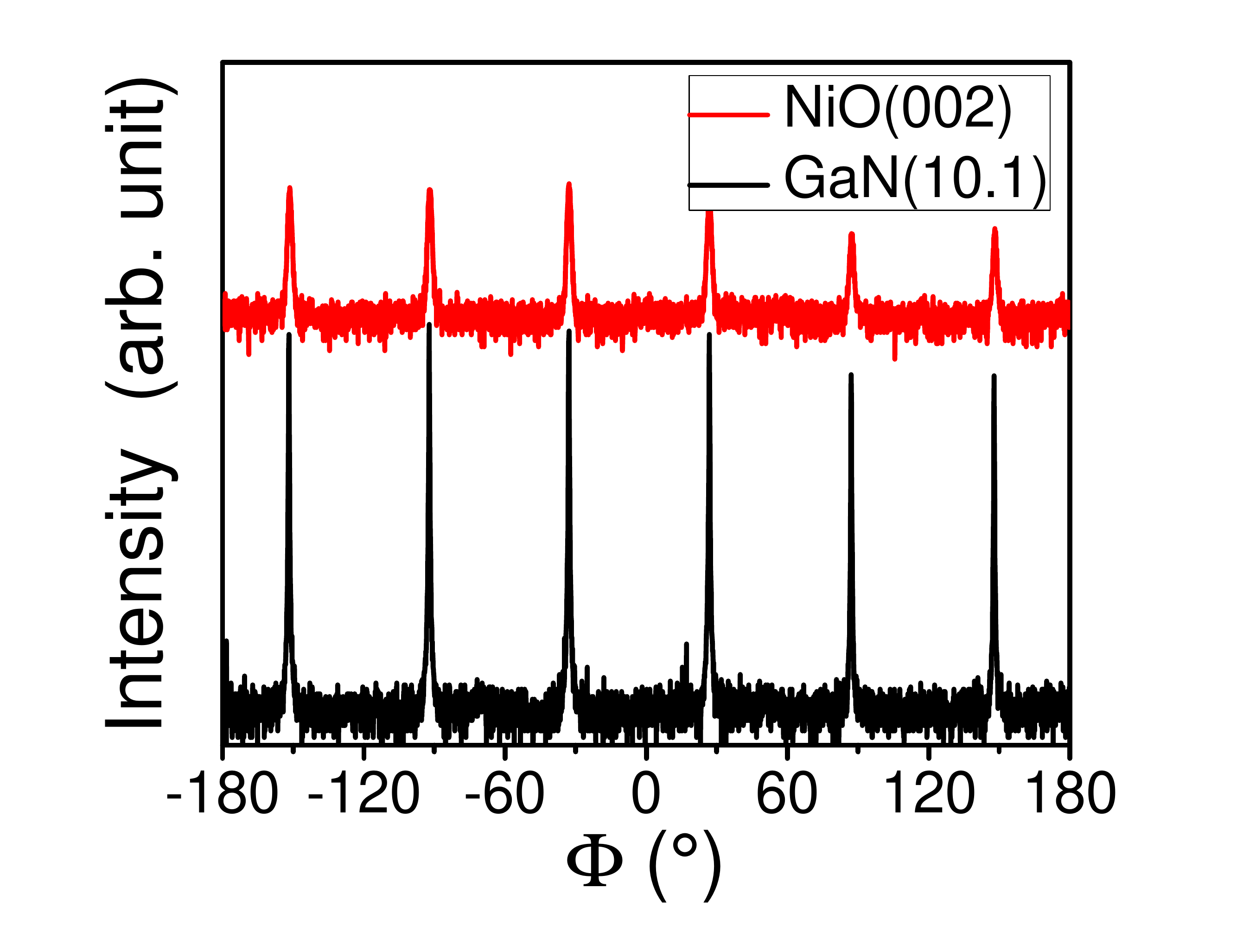}
\par\end{centering}
\caption{XRD \textgreek{F}\protect\nobreakdash-scan for S500 of tilted lattice
planes as indicated in the legend. \label{fig:Phi Scan}}
\end{figure}

\noindent Distinct Laue oscillations are visible, which suggest a
sufficiently smooth interface and surface for all samples. However,
for S850, a strong reduction of the oscillations indicates a rougher
interface at this growth temperature. Compared to the (111) reflex
of bulk NiO ($a_{NiO,}^{bulk}=\mathrm{0.4176}$~nm), the NiO(111)
peak of the S100 sample is shifted to lower angles, which indicates
tensile strain with a lattice constant around 0.419\,nm. Compressive
out\nobreakdash-of\nobreakdash-plane strain is expected due to the
Poisson effect and the higher lattice constant of the substrate surface
lattice. The higher lattice constant for S100 could arise for example
from defects or the low grain size (see below). The peaks of the samples
grown at higher temperatures are shifted to higher angles, indicative
of compressive strain with an out\nobreakdash-of\nobreakdash-plane
lattice constant down to 0.416\,nm. The in-plane epitaxial relationship
of the GaN/NiO samples was determined by \textgreek{F}\nobreakdash-scans
of the two skew symmetric reflections GaN(10.1) and NiO(002). The
result for S500, representing all samples, is shown in Fig.~\ref{fig:Phi Scan}.
The GaN(10.1) reflection shows a sixfold rotational symmetry as expected
from the hexagonal wurtzite-structure. However, the NiO(002) peak
also shows a sixfold rotational symmetry despite the expected threefold
one of the (111)-oriented rock~salt structure.\citep{Domain_Summary_2}
This observation indicates an epitaxial growth with the formation
of two domains rotated by 60$^{\circ}$ with respect to each other. Similar
intensities of the two domains indicate a homogeneous distribution
of both.

\noindent As the peaks of layer and substrate are at the same \textgreek{F}
angles, the following in-plane epitaxial relationship of both NiO
domains, with respect to the GaN substrate, can be defined: 
\begin{equation}
\begin{array}{ccc}
 & \mathrm{NiO(1\overline{1}0)\:||\:GaN(11.0)}\end{array}\label{eq:Grain1}
\end{equation}

\noindent \begin{center}
for domain 1 and
\begin{equation}
\begin{array}{ccc}
 & \mathrm{NiO(10\overline{1})\:||\:GaN(11.0)}\end{array}\label{eq:Grain2}
\end{equation}
\par\end{center}

\noindent \begin{center}
for domain 2.
\par\end{center}

\begin{figure}[th]
\noindent \begin{centering}
\includegraphics[width=7cm]{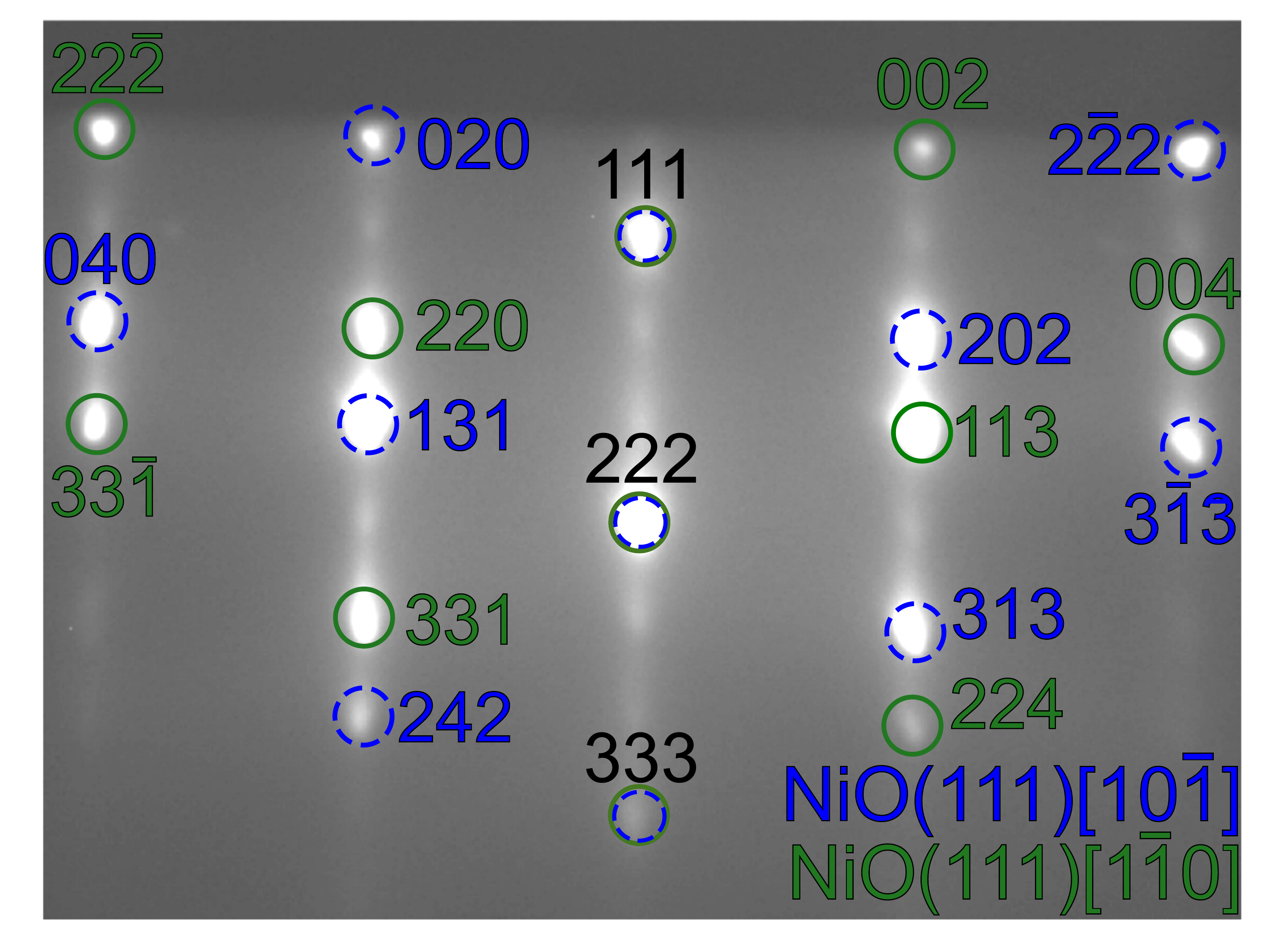}
\par\end{centering}
\caption{RHEED image of a GaN/NiO sample. The domains for the different diffraction
spots are defined using the diffraction pattern simulation of \textquotedbl Web\protect\nobreakdash-EMAPS\textquotedbl .\citep{WebEMAPS}
The patterns also indicate a two domain growth of NiO and are marked
by green and blue circles.\label{fig:Rheed Pattern}}
\end{figure}

\noindent In fact, the two domains could already be identified during
growth by the spotty RHEED patterns arising from simultaneous transmission
diffraction of the electron beam through the asperities of multiple
domains. The superposition of the diffraction spots of the two domains,
taken along the GaN(11.0) azimutal direction, are shown in Fig.~\ref{fig:Rheed Pattern}.
Blue and green circles mark the expected diffraction spots of the
different domains as determined by simulation of electron transmission
diffraction for the two domain orientations determined by the XRD
measurements. Their coincidence with the experimental spots in the
RHEED image confirms that RHEED can be used as an in-situ tool to
determine domain orientations.

\begin{figure}[h]
\noindent \begin{centering}
\includegraphics[width=8.5cm]{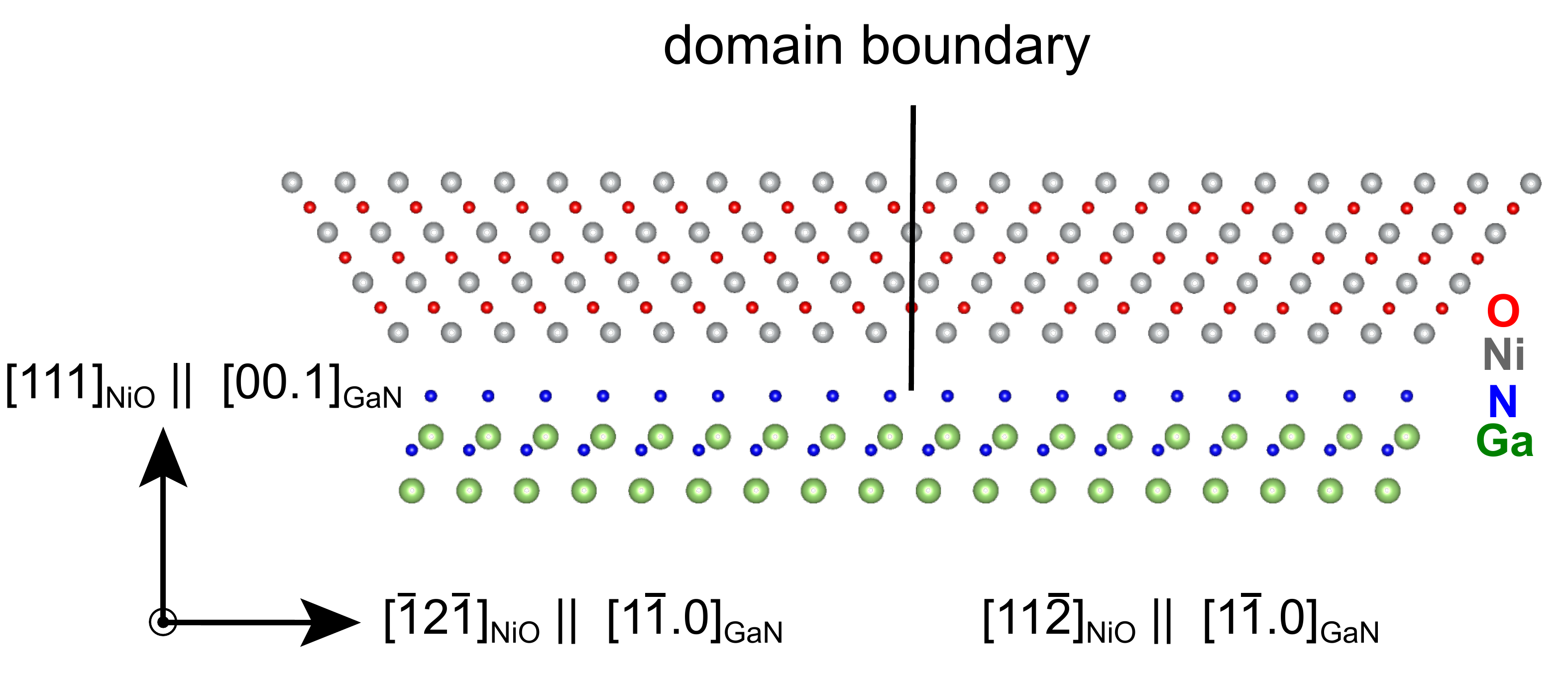}
\par\end{centering}
\caption{Diagram of the growth of NiO on GaN by two domains. The resulting
domain boundary is marked. The atoms are marked by colors where O
is red, Ni is \foreignlanguage{british}{grey}, N is blue and gallium
is green.\label{fig:Domain Model}}
\end{figure}

\noindent A schematic cross section illustrating the two domains of
NiO on GaN can be seen in Fig.~\ref{fig:Domain Model}. For the diagram
a N\nobreakdash-terminated GaN surface was assumed, since the NiO
layer should start with Ni due to the oxygen free start of our growth
process. This growth of two rotational domains is expected as a result
of the different rotational symmetries of substrate (sixfold) and
layer (threefold),\citep{Domain_Summary,Domain_Summary_2} and has
also been observed for the growth of MgO(111) and In$_{2}$O$_{3}$(111)
on GaN(00.1) by Craft et al.\citep{MgO_on_GaN} and Tsai et al.,\citep{Tsai2015},
for MgO(111) on ZnO(00.1) by Saitoh et al.\citep{MgO_on_ZnO} or for
NiO(111) on Al$_{2}$O$_{3}$(00.1) by Lee et al.\citep{NiO_on_Saph}

\noindent The theoretical in-plane mismatch between NiO and GaN can
be derived from the basis vectors $\mathrm{a_{Ni}=2.95~\mathring{A}}$
and $\mathrm{a_{N}=3.19~\mathring{A}}$ of the NiO and GaN surface
unit cells. Thus, the in-plane strain for lattice matching epitaxy
(LME) calculated by:\citep{DomainEpitaxy}
\begin{equation}
\epsilon_{\textrm{LME}}=\frac{a_{\mathrm{substrate}}}{a_{\mathrm{layer}}}-1\label{eq:epsi_LME}
\end{equation}

\noindent would result in a tensile strain of +8.1~\%, which is in
the critical range (7-8~\%) for LME.\citep{DomainEpitaxy} LME describes
a one-to-one matching of lattices in pseudomorphic growth. At a critical
thickness, dislocations formed at the growth surface can glide to
the interface to relax the layer. For a higher mismatch (>7-8\,\%),
the growth mechanism ``domain matching epitaxy'' (DME) has been
proposed by Narayan and Larson,\citep{DomainEpitaxy} which describes
the growth by coinciding super cells (SC) of the surface unit cells
of film and substrate. SCs are created by multiplying the surface
unit cell (SUC) to create a better matching. For DME the residual
strain is calculated by:\citep{DomainEpitaxy}
\begin{equation}
\epsilon_{\textrm{DME}}=\frac{n\cdot\left|\boldsymbol{a}_{\textrm{substrate}}\right|}{m\cdot\left|\boldsymbol{a}_{\textrm{layer}}\right|}-1\label{eq:epsi_DME}
\end{equation}

\begin{figure}[h]
\noindent \begin{centering}
\includegraphics[width=6.5cm]{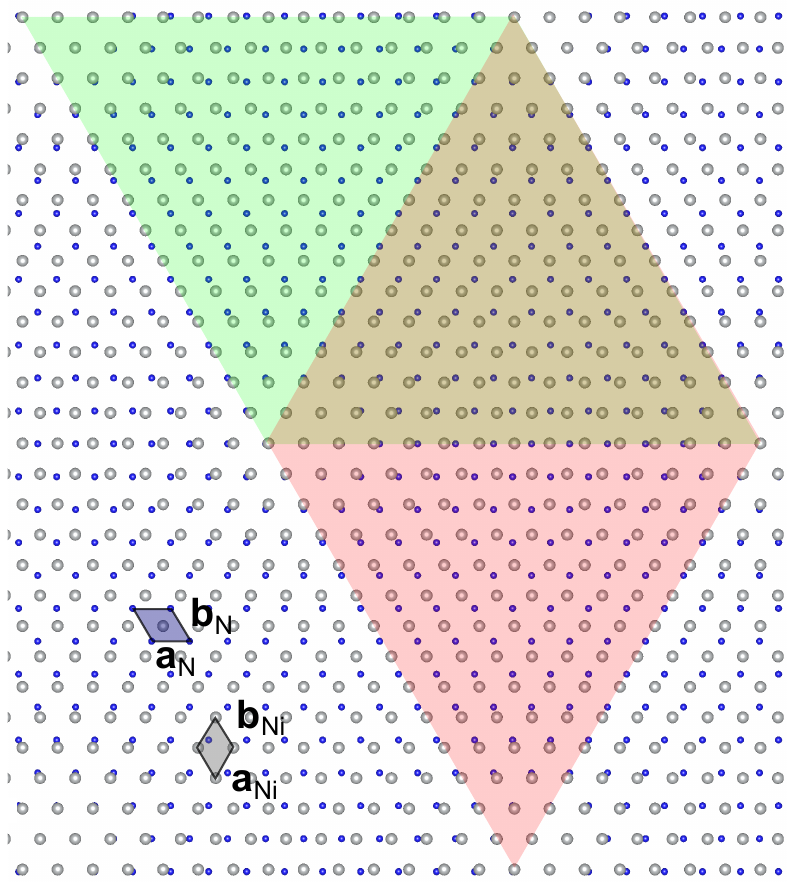}
\par\end{centering}
\caption{In-plane schematic of the two possible super cells to achieve a misfit
strain of only 0.41\,\% between GaN and NiO by domain matching epitaxy.
Ni atoms are marked in grey and the N atoms in blue. The same colouring
is used for the SUCs.\label{fig:in-plane-schematic-SUC}}
\end{figure}

\noindent Here, $\mathrm{m}$ is the number of Ni SUCs, which are
needed for the SC, and $\mathrm{n}$ is the number of needed N SUCs.
(For a similar format as Eq.~\ref{eq:epsi_LME}, Eq.~\ref{eq:epsi_DME}
was slightly changed compared to Ref.~{[}\citen{DomainEpitaxy}{]}.)
The lowest residual tensile strain between NiO and GaN can be achieved
for $\mathrm{m}$=14 and $\mathrm{n}$=13, leading to a strain of
0.41\,\%. Thus, a super cell is approximately 4\,nm large and can
be described by:
\[
\begin{array}{ccc}
\mathrm{a_{SC,\,N}=13\cdot a_{N}} &  & \mathrm{a_{SC,\,Ni}=14\cdot a_{Ni}}\\
\mathrm{b_{SC,\,N}=13\cdot b_{N}} &  & \mathrm{b_{SC,\,Ni}=14\cdot b_{Ni}}
\end{array}
\]
Here, $\mathrm{a_{SC}}$ and $\mathrm{b_{SC}}$ are the vectors of
the created SC. An in-plane schematic of the created SC ($\mathrm{m}$=14,
$\mathrm{n}$=13) is shown in Fig.~\ref{fig:in-plane-schematic-SUC}~(green).
The same SC can be created by a 60$^{\circ}$ rotated Ni SUC, which only leads
to a small difference in the definition of the SC but creates the
same misfit of 0.41\,\% (red). The advantage of DME growth is that
the misfit dislocation network is established so early in the growth
that no threading arms are generated in the layer above. Layers grown
by LME, in contrast, typically possess threading arms of the disclocation
halfloops needed for establishing misfit dislocation segments at the
interface.\citep{DomainEpitaxy}

\subsubsection*{Transmission electron microscopy}

\begin{figure}
\noindent \begin{centering}
\includegraphics[width=8cm]{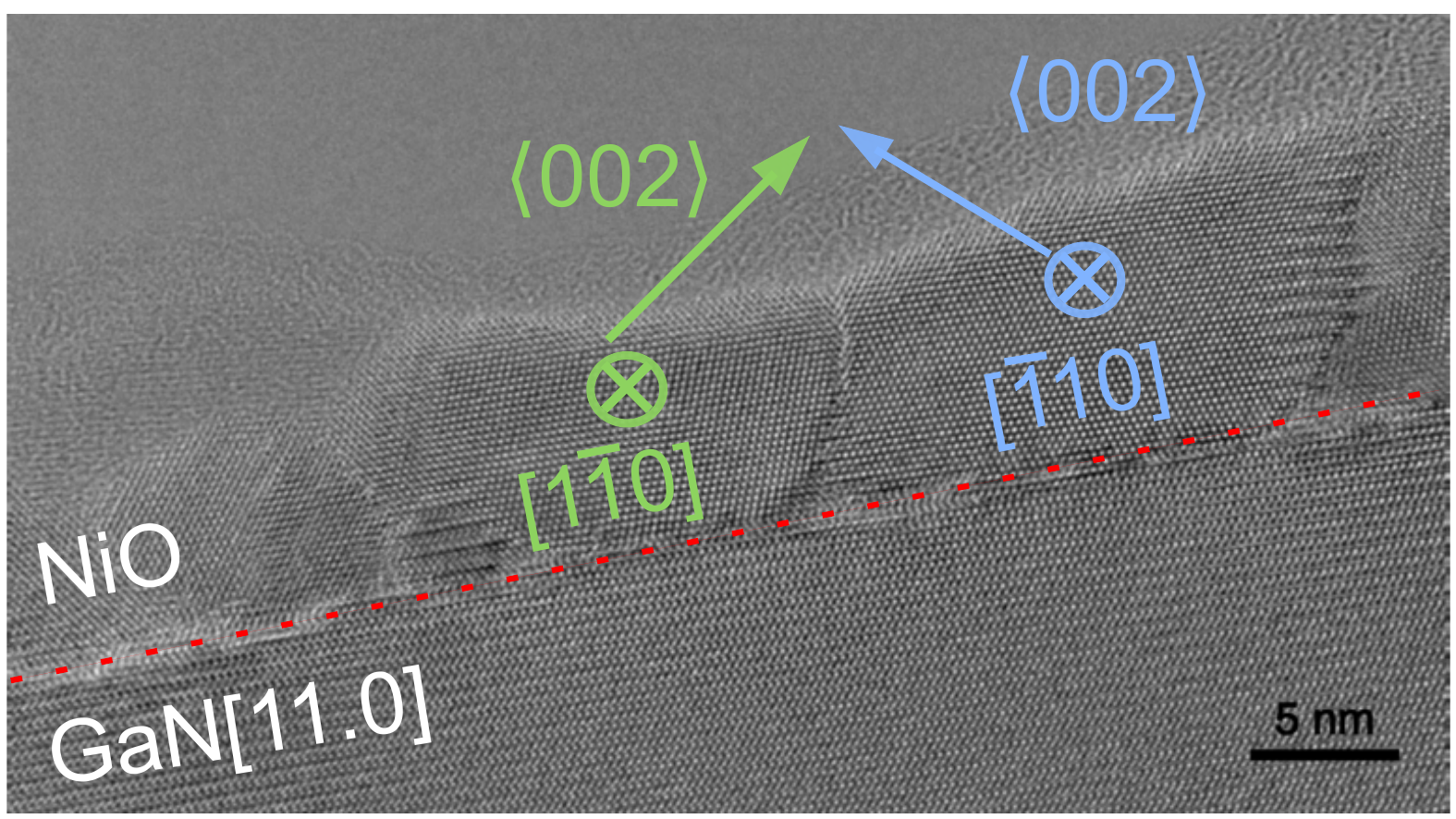}
\par\end{centering}
\caption{Cross-sectional TEM image of S700 showing the two rotational $\mathrm{NiO\left\langle 110\right\rangle \parallel GaN\left[11.0\right]}$
domains. The different rotations are indicated by the different $\left\langle 002\right\rangle $
directions. A red dashed line indicates the interface between GaN
and NiO. \label{fig:TEM-image-domains}}
\end{figure}

Cross-sectional TEM of S100, S700, and S850 is used to investigate
the possible creation of interfacial oxides, the grain sizes and the
epitaxial relationship between substrates and layers. Mainly, columnar
grains with $\mathrm{NiO\left\langle 1\bar{\mathrm{{1}}}0\right\rangle \parallel GaN\left[11.0\right]}$
are found for all growth temperatures, confirming the orientations
observed in XRD (see Eq.\ \ref{eq:Grain1} and Eq.~\ref{eq:Grain2})
and RHEED. The existence of the two different domains is confirmed
by the observation of mirrored $\mathrm{\left[002\right]}$ directions
for the grains illustrated by arrows in Fig.~\ref{fig:TEM-image-domains}.
 Occasionally, also $\mathrm{\mathrm{NiO\left\langle 11\bar{\mathrm{{2}}}\right\rangle \parallel GaN\left[11.0\right]}}$
grains can be seen in the TEM images. This orientation is still consistent
with NiO(111) growth and a rotation angle of 30$^{\circ}$ with respect to the
other grains. The absence of any NiO(002) reflections between the
substrate reflections in the \textgreek{F}\nobreakdash-scan (cf.
Fig.~\ref{fig:Phi Scan}), however, suggests only a minor fraction
of those 30$^{\circ}$ rotated NiO grains.

\subsubsection*{Electron backscatter diffraction}

Alternatively, EBSD measurements confirm the (111) out-of-plane direction
of the NiO film and the presence of twin domains rotated by 60$^{\circ}$. The
probing depth of EBSD is $\mathrm{\approx}$20~nm, and, for thinner
films, the GaN Kikuchi pattern is superimposed on the NiO pattern.
Therefore, we chose the thickest NiO layer S300 (d$\mathrm{_{XRD}}$=
20.9~nm) for the EBSD investigation (cf. Tab.~\ref{tab:size+d-of-AFM+XRD}).
Examples of Kikuchi patterns recorded for the two twin orientations
are given in Fig.~\ref{fig:EBSD}. The Kikuchi bands used to index
the pattern are marked by coloured lines. As expected for crystal
twins, both patterns share several of the main bands, but also contain
distinctive bands indicative of the respective in-plane orientation.
Nevertheless, each of the patterns also contains weak bands from the
opposite direction as highlighted by the dashed lines in Fig.~\ref{fig:EBSD}.
In fact, the domain size derived from the transmission electron micrographs
(see Tab.~\ref{tab:size+d-of-AFM+XRD}) between 10 and 25~nm, depending
on the growth temperature, is on the order of the lateral resolution
of the EBSD measurements, which explains the overlapping of Kikuchi
patterns. A further consequence of this small grain size is that maps
of the in-plane orientation obtained from an automatic indexing of
the Kikuchi patterns recorded when stepping the electron beam over
the sample surface lead to a misrepresentation of the domain size
as discussed in the supplementary material.
\begin{figure}
\noindent \begin{centering}
\includegraphics[width=8cm]{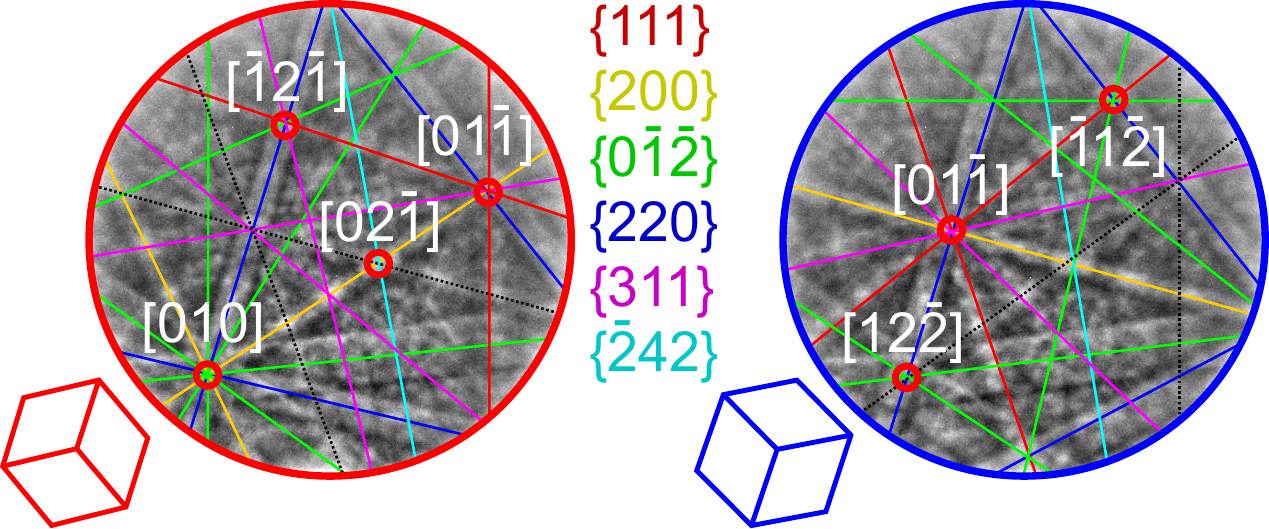}
\par\end{centering}
\caption{Kikuchi patterns for the two twin orientations of cubic NiO found
in the layers accompanied by a wireframe representation of the indexed
orientation. On top of the patterns, the bands used during the indexing
are highlighted by colored lines, where each color corresponds to
one reflector (strongly reflecting plane). Additionally, major zone
axis directions (intersection points of bands) are labeled. Note that
both patterns also contain some weaker bands that correspond to the
other twin direction as highlighted by the black dashed lines.\label{fig:EBSD}}
\end{figure}

\subsection{Strain\label{subsec:Strain}}

\noindent The in-plane lattice parameter of the NiO films in S100--S700,
measured by grazing incidence XRD at room temperature, was used to
calculate the in-plane strain $\varepsilon_{||}\mathrm{(RT)}$ in
these films. The results are shown in Tab.~\ref{tab:Thermal-strain}.
Besides the in\nobreakdash-plane strain at growth temperature $\mathrm{\varepsilon_{||}}\mathrm{(T_{g})}$,
the calculated strain also includes the thermal strain $\mathrm{\varepsilon_{T}}$
arising during cooldown to room temperature after sample growth as
a result of the different thermal expansion of GaN and NiO ($\varepsilon_{||}\mathrm{(RT)}=\mathrm{\varepsilon_{||}}\mathrm{(T_{g})}+\mathrm{\varepsilon_{T}}$).
Thermal strain can be estimated from the thermal expansion coefficients
for the two materials over the temperature range, resulting in $\alpha_{\mathrm{NiO}}=$7.93$\mathrm{\cdot10^{-6}}$\,$\mathrm{\frac{1}{K}}$
for NiO\citep{Thermal_NiO} and $\alpha_{\mathrm{GaN}}=$5.59$\mathrm{\cdot10^{-6}}$\,$\mathrm{\frac{1}{K}}$
for GaN.\citep{thermal_GaN} The thermal strain depends on the temperature
difference ($\mathrm{\bigtriangleup T}$) between growth temperature
and room temperature: 
\begin{equation}
\varepsilon_{T}=(\alpha_{\mathrm{GaN}}-\alpha_{\mathrm{NiO}})\cdot\bigtriangleup T\label{eq:thermal strain}
\end{equation}

\noindent The resulting values for $\mathrm{\bigtriangleup T}$ and
$\mathrm{\varepsilon_{T}}$ can be found in Tab.~\ref{tab:Thermal-strain}.
Sample S850 was excluded from strain evalutation, since the decomposition
of the GaN substrate and the interfacial oxide (see section~\ref{subsec:Interfacial-oxide})
prevents a meaningful comparison with the other samples. Samples grown
between 300~$^{\circ}$C and 700~$^{\circ}$C show a similar strain $\varepsilon_{||}\mathrm{(T_{g})}$
for their growth temperature. The values are comparable with the residual
strain by DME growth of 0.41~\%. However, since relaxation is a gradual
process, which can lead to small residual strain even for thicknesses
above the critical thickness, LME can not be excluded. S100 shows
a much lower in\nobreakdash-plane strain, indicating an almost relaxed
layer. This could be created by a change in the relaxed lattice constant
of the layer due to the higher defect density, as it was already mentioned
for the out\nobreakdash-of\nobreakdash-plane strain (see section~\ref{subsec:Epitaxial-relationship})
and will be further discussed within the Raman results.

\noindent 
\begin{table}
\caption{Temperature difference ($\mathrm{\bigtriangleup T}$), calculated
thermal ($\mathrm{\varepsilon_{T}}$) and measured in\protect\nobreakdash-plane
($\mathrm{\mathrm{\varepsilon_{\parallel}}}$) strain for NiO grown
on GaN for all growth temperatures.\label{tab:Thermal-strain}}

\centering{}%
\begin{tabular*}{8cm}{@{\extracolsep{\fill}}>{\centering}m{1.5cm}>{\centering}m{1.5cm}>{\centering}m{1.5cm}>{\centering}m{1.5cm}>{\centering}m{1.5cm}}
\toprule 
\textbf{$\mathbf{\mathbf{T}_{\mathbf{g}}}$ }

\textbf{{[}$^{\circ}$C{]}} & \textbf{$\mathbf{\boldsymbol{\mathrm{\mathbf{\mathrm{\varepsilon_{||}}}}\mathrm{(RT)}}}$}

\textbf{{[}\%{]}} & \textbf{$\mathrm{\mathbf{\mathrm{\mathbf{\bigtriangleup T}}}}$}

\textbf{{[}K{]}} & \textbf{$\mathbf{\boldsymbol{\mathrm{\mathbf{\mathrm{\varepsilon_{T}}}}}}$}

\textbf{{[}\%{]}} & \textbf{$\mathrm{\mathbf{\boldsymbol{\mathrm{\varepsilon_{||}}\mathrm{(T_{g})}}}}$}

\textbf{{[}\%{]}}\tabularnewline
\midrule 
\textbf{100} & 0.06 & -75 & 0.02 & 0.04\tabularnewline
\midrule 
\textbf{300} & 0.59 & -275 & 0.06 & 0.53\tabularnewline
\midrule 
\textbf{500} & 0.42 & -475 & 0.11 & 0.31\tabularnewline
\midrule 
\textbf{700} & 0.47 & -675 & 0.16 & 0.31\tabularnewline
\bottomrule
\end{tabular*}
\end{table}

\begin{figure}
\begin{centering}
\includegraphics[width=8cm]{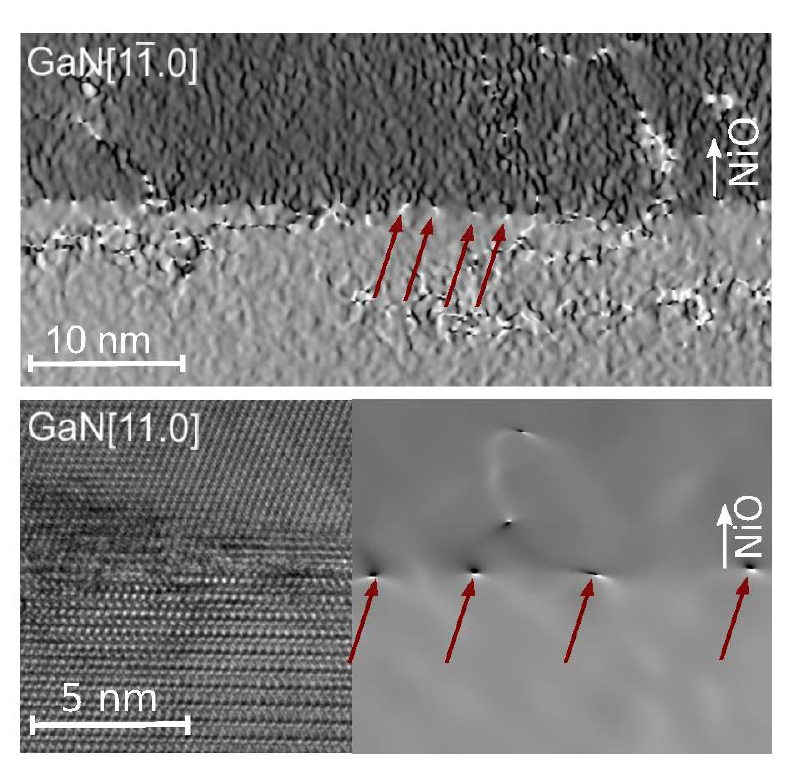}
\par\end{centering}
\caption{Strain map of S700 derived from an HRTEM image in the GaN{[}1$\bar{\mathrm{1}}$.0{]}
and GaN{[}11.0{]} projection is shown, as well as the corresponding
HRTEM image from the GaN{[}11.0{]} projection. Misfit disclocations
are visible every 2 or 3\,nm (red arrows), for the GaN{[}1$\bar{\mathrm{1}}$.0{]}
and GaN{[}11.0{]}, respectively, as expected for a relaxed NiO layer.
A similar map can be found for S100 in the GaN{[}11.0{]} direction
(see Fig. S2).\label{fig:Strain-map}}
\end{figure}

\noindent From HRTEM images, the strain can be estimated by the geometric
phase analysis.\citep{Hytch_1998} In particular, dislocations have
a characteristic strong and localized bright-dark contrast at their
cores originating from the change from high compressive to high tensile
strain. Taking a look at the strain map of S700 by TEM, a clear misfit
dislocation network can be seen at the interface (see Fig.~\ref{fig:Strain-map}),
indicating an almost relaxed NiO layer which appears darker due to
the narrower spacing of the NiO lattice compared to GaN used as reference.
The sample S100 shows the same network (see Fig. S2). In the GaN $[1\bar{1}.0]$
projection, every 13th GaN(11.0) plane an additional NiO(110) plane
is expected in relaxed NiO. Therefore, about every 2\,nm, a dislocation
contrast in the geometric phase analysis of the HRTEM pattern occurs.
While the GaN(1$\bar{\mathrm{1}}$.0) planes are parallel to the NiO(110)
planes, there is no low index lattice plane in the NiO layer parallel
to the GaN(11.0) planes. Instead, one can see a continuation of GaN$\mathrm{\{1\bar{{1}}.1\}}$
planes by NiO$\mathrm{\{111\}}$ and NiO\{002\} planes with an inclination
of around 6.5$^{\circ}$ for relaxed NiO(111) on GaN(00.1). In this case, NiO
will have an additional (111) plane compared to the number of GaN$\mathrm{\{1\bar{{1}}.1\}}$
planes around every 3\,nm, when viewed in the GaN{[}11.0{]} direction,
respectively for the NiO(002) plane. Again, the strain analysis shows
dislocation contrasts with this spacing at the interface. One should
mention that for such narrowspaced misfit dislocation networks the
characterization of the dislocation types is ambigiuous in wurtzite
materials.\citep{Zhang:2012,Kioseoglou:2012} Especially with dislocations
running transversely to the projection direction causing a distortion
of the high resolution pattern at the interface, which hinders a direct
interpretation of the pattern in our case. However, the dislocation
network can either occur due to relaxation above the critical thickness
or as a result of the growth by DME. The small domain sizes make it
impossible to differentiate between the two modes by analysing the
threading dislocation density.

\noindent 

\subsection{Surface morphology}

\begin{figure}[h]
\begin{centering}
\includegraphics[width=7.5cm]{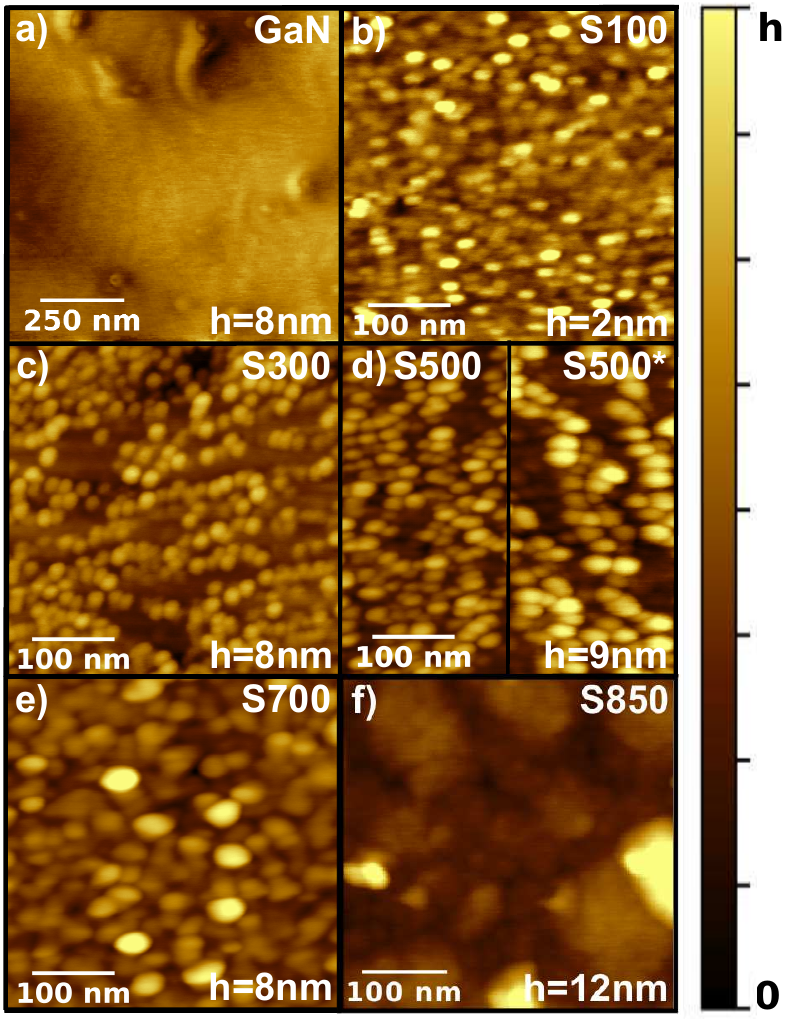}
\par\end{centering}
\caption{AFM images of the GaN template, and the NiO layers grown at different
temperatures. The main difference is the lateral island size. For
S100--S500 the islands are smaller compared to those of S700. A drastic
increase is found for S850. \label{fig:AFM-images}}
\end{figure}

AFM images of all samples and a clean GaN substrate as reference are
shown in Fig.~\ref{fig:AFM-images}. The AFM images of the NiO layers
show a morphology defined by islands. Their lateral size (\textbf{\small{}$S\mathrm{_{AFM}}$}),
evaluated with the program 'Gwyddion' by taking the average lateral
size of ten islands, can be found in Tab.~\ref{tab:size+d-of-AFM+XRD}
together with the layer thicknesses measured by XRD ($d\mathrm{_{XRD}}$).
In addition, the root-mean-square roughness (\textbf{\small{}$R\mathrm{_{RMS}}$}),
determined for 0.4$\mathrm{\times}$0.4\,$\text{\textmu}\mathrm{m^{2}}$
images, is given. The film roughness of $\mathrm{\approx1\thinspace}$nm
for S100--S700 is larger than that of the GaN substrate ($\mathrm{\approx}$0.3~nm).
A clear difference in the topology can be seen for S850, showing drastically
enlarged islands. 

\noindent Film thickness and island size were also extracted from
cross-sectional TEM images as shown in Tab.~\ref{tab:TEM-size-d}.
The TEM image of S700 (see Fig.~\ref{fig:TEM-image-domains}) clearly
shows a columnar domain growth of the layer, correlating the formation
of islands with the formation of the two rotational domains. The thicknes
$d\mathrm{_{TEM}}$ is about 2.5\,nm thicker than the thickness $d\mathrm{_{XRD}}$
calculated from the Laue oscillation seen in the $\mathrm{2\theta-\omega}$
scans (see Fig.~\ref{fig:2Theta Combination}). However, $d\mathrm{_{XRD}}$
depends strongly on the crystal quality and includes planes with similar
distances, which often leads to an underestimation of the thickness
(e.g. strained interface region). The thickness of S850 could not
be measured by XRD as a result of the reduced oscillation intensity.
The TEM images show a variation of the thickness by $\mathrm{\pm}$2\,nm
over the sample, due to roughness of surface and interface. 
\begin{table}
\caption{Calculated layer thicknesses from oscillation fringes in XRD on-axis
$\mathrm{2\theta-\omega}$ scans (d$\mathrm{_{XRD}}$), average island
sizes (\textbf{\small{}$S\mathrm{_{AFM}}$}) and roughness (\textbf{\small{}$R\mathrm{_{RMS}}$})
measured from AFM images. Results for NiO layer thickness ($d\mathrm{_{TEM}}$)
and island size ($S\mathrm{_{TEM}}$) from TEM images in comparison
for all growth temperatures. In addition, full width at half maximum
($\mathrm{\triangle\omega}$) from a Gaussian fit for all temperatures
measured at the NiO(111) XRD peak. \label{tab:TEM-size-d}\label{tab:size+d-of-AFM+XRD}}

\noindent \centering{}%
\begin{tabular*}{8.5cm}{@{\extracolsep{\fill}}>{\centering}m{0.9cm}>{\centering}m{1.12cm}>{\centering}m{1.16cm}>{\centering}m{1.16cm}>{\centering}m{1.18cm}>{\centering}m{1.18cm}>{\centering}m{1.1cm}}
\toprule 
\textbf{$\mathbf{\mathrm{\mathrm{\mathbf{T}_{\mathbf{g}}}}}$ {[}$^{\circ}$C{]}} & \textbf{\small{}$\mathbf{d}\mathrm{_{\mathbf{XRD}}}$}\textbf{ {[}nm{]}} & \textbf{\small{}$\mathbf{d}\mathrm{\mathbf{_{TEM}}}$}\textbf{ {[}nm{]}} & \textbf{\small{}$\mathbf{S}\mathrm{\mathbf{_{AFM}}}$}\textbf{ {[}nm{]}} & \textbf{\small{}$\mathbf{S}\mathrm{\mathbf{_{TEM}}}$}\textbf{ {[}nm{]}} & \textbf{\small{}$\mathbf{R}\mathrm{\mathbf{_{RMS}}}$}\textbf{ {[}nm{]}} & \textbf{$\mathbf{\boldsymbol{\mathrm{\triangle\omega}}}$ {[}$^{\circ}${]}}\tabularnewline
\midrule 
\textbf{100} & 17.0 & 19.5 & 35 & 10-20 & 0.35 & 0.153\tabularnewline
\midrule 
\textbf{300} & 20.9 &  & 30 &  & 0.93 & 0.165\tabularnewline
\midrule 
\textbf{500} & 17.2 &  & 35 &  & 1.30 & 0.166\tabularnewline
\midrule 
\textbf{500{*}} & 25.2 & 25.0 & 50 & 10-30 & 2.02 & 0.152\tabularnewline
\midrule 
\textbf{700} & 17.4 & 20.0 & 45 & 20-25 & 1.11 & 0.163\tabularnewline
\midrule 
\textbf{850} &  & 15.0 & 165 & $\sim$120 & 2.38 & 0.722\tabularnewline
\bottomrule
\end{tabular*}
\end{table}

\noindent The domain sizes measured by TEM are much smaller than the
islands measured by AFM, likely related to the convolution with the
tip shape and morphology. Therefore, the AFM can not resolve all trenches
between domains which leads to the apparent larger island size than
domain size. However, AFM (surface islands) and TEM (domains) measurements,
show a similar trend for both methods, i.e. an increase of the domain/island
size with increasing growth temperature. The sample S850 has the biggest
domains (cf. Tab.~\ref{tab:TEM-size-d}). The increasing domain size
indicates a higher adatom surface diffusion length consistent with
a higher growth temperature, as it was also observed for the growth
on MgO(100).\citep{MgO_NiO} 

\subsection{Raman quality metrics\label{subsec:Raman-quality-metrics}}

\begin{figure}
\noindent \begin{centering}
\includegraphics[width=8.5cm]{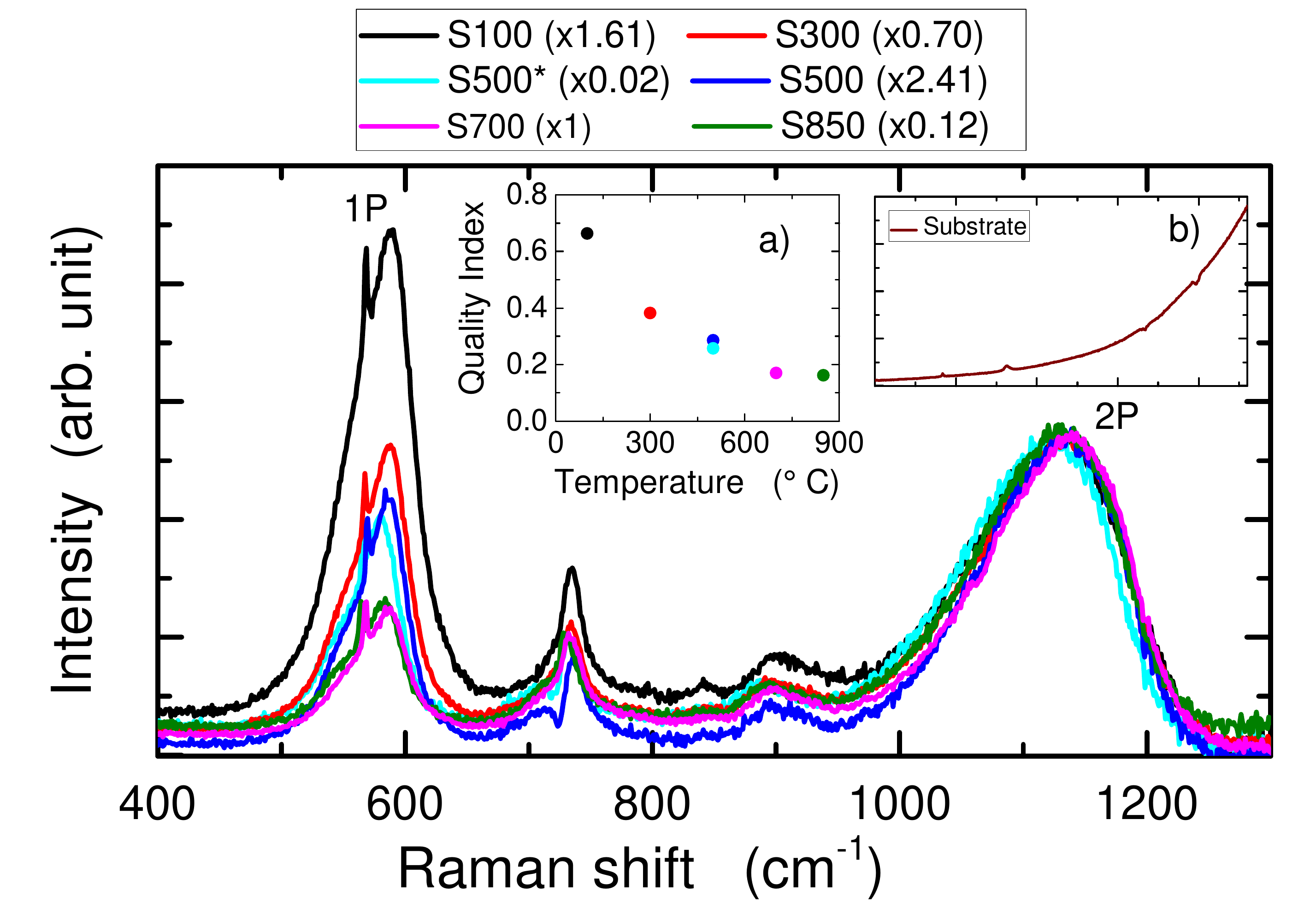}
\par\end{centering}
\caption{Raman spectra of all NiO layers excited at 3.81 eV. A featureless
background (high-energy tail of the near bandgap photoluminescence
from the GaN substrate) has been subtracted (see example in inset
(b)). The quality index Q as a function of the growth temperature
is shown in inset (a). \label{fig:Raman}}
\end{figure}

\noindent Figure~\ref{fig:Raman} displays room temperature Raman
spectra of all NiO layers excited at 3.81 eV, nearly resonant with
the bandgap of NiO. The spectra exhibit peaks due to first\nobreakdash-order
(580~cm$^{\mathrm{-1}}$) and second\nobreakdash-order (730, 900,
and 1130~cm$^{\mathrm{-1}}$) scattering by optical phonons in NiO.
Superimposed on these spectral features, the E$_{2}$ (high) and A$_{1}$
(LO) Raman peaks of the GaN substrate are detected at 570 and 735~cm$^{\mathrm{-1}}$,
respectively.\citep{Harima_GaN} Since first-order Raman scattering
by optical phonons is forbidden for the the rock salt structure, the
ratio between the intensities of the first-order peak at 580~cm$^{\mathrm{-1}}$
(1P) and the second-order peak at 1130~cm$^{\mathrm{-1}}$ (2P) from
NiO has been established as a quality metrics (Q) for NiO films on
MgO(100):\citep{MgO_NiO}

\noindent 
\begin{equation}
Q=\frac{I_{\mathrm{1P}}}{I_{\mathrm{2P}}}\label{eq:Q}
\end{equation}

\noindent A perfect crystal is characterized by a quality index (Q)
of zero, since $I\mathrm{_{1P}}$ = 0. The value of Q increases with
the density of crystal defects. In the case of NiO, a potentially
desired defect could be Ni vacancies to induce $\mathit{p}$-type
conductivity.\citep{Dietz_Raman} Since our layers are always insulating
as shown on MgO,\citep{MgO_NiO} the increase in the 1P peak cannot
be solely related to Ni vacancies.

\noindent Applying this metrics to the NiO layers grown on GaN, we
find a reduction of Q with increasing growth temperature (cf. inset
in Fig.~\ref{fig:Raman}) indicating a decrease in the density of
crystal defects. This trend is consistent with an increasing domain
size, meaning a reduction of grain boundaries (see TEM images), and
an increased adatom surface diffusion length with the increasing growth
temperature. The quality is further increased by a reduction of crystal
defects and the formation of a thermodynamically stable structure
using a higher growth temperature. Although the domain size of S850
is relatively large, its quality index is almost the same as that
of S700. Probably, the quality of S850 is reduced by the interfacial
oxide grown under those conditions (see section~\ref{subsec:Interfacial-oxide}).
This influence can be seen in the FWHM measured by a rocking curve
of the NiO(111) peak. S100 - S700 show values around 0.16$^{\circ}$, S850 however,
shows a $\mathrm{\triangle\omega}$ of 0.72$^{\circ}$. The results and different
trends from the two methods show again, the higher sensitivity of
Raman on the crystal imperfection of a thin layer. As already observed
in Ref.~{[}\citen{MgO_NiO}{]}, XRD ($\triangle\omega$) is mainly
influenced by the interface structure. Compared to the Q of NiO on
MgO(100)~(<0.2) and FWHM~(0.05$^{\circ}$-0.07$^{\circ}$) grown under the same conditions,
the values are higher for NiO on GaN, which is likely related to the
formation of rotational domain grain boundaries on the hexagonal GaN.
In the case of the cubic MgO(100) substrate, no domains have been
observed.\citep{MgO_NiO} Nevertheless, a high temperature is again
beneficial for the growth of a high quality layer.

\subsection{Interfacial oxide and Ni preflow\label{subsec:Interfacial-oxide}}

\begin{figure}
\noindent \begin{centering}
\includegraphics[width=8cm]{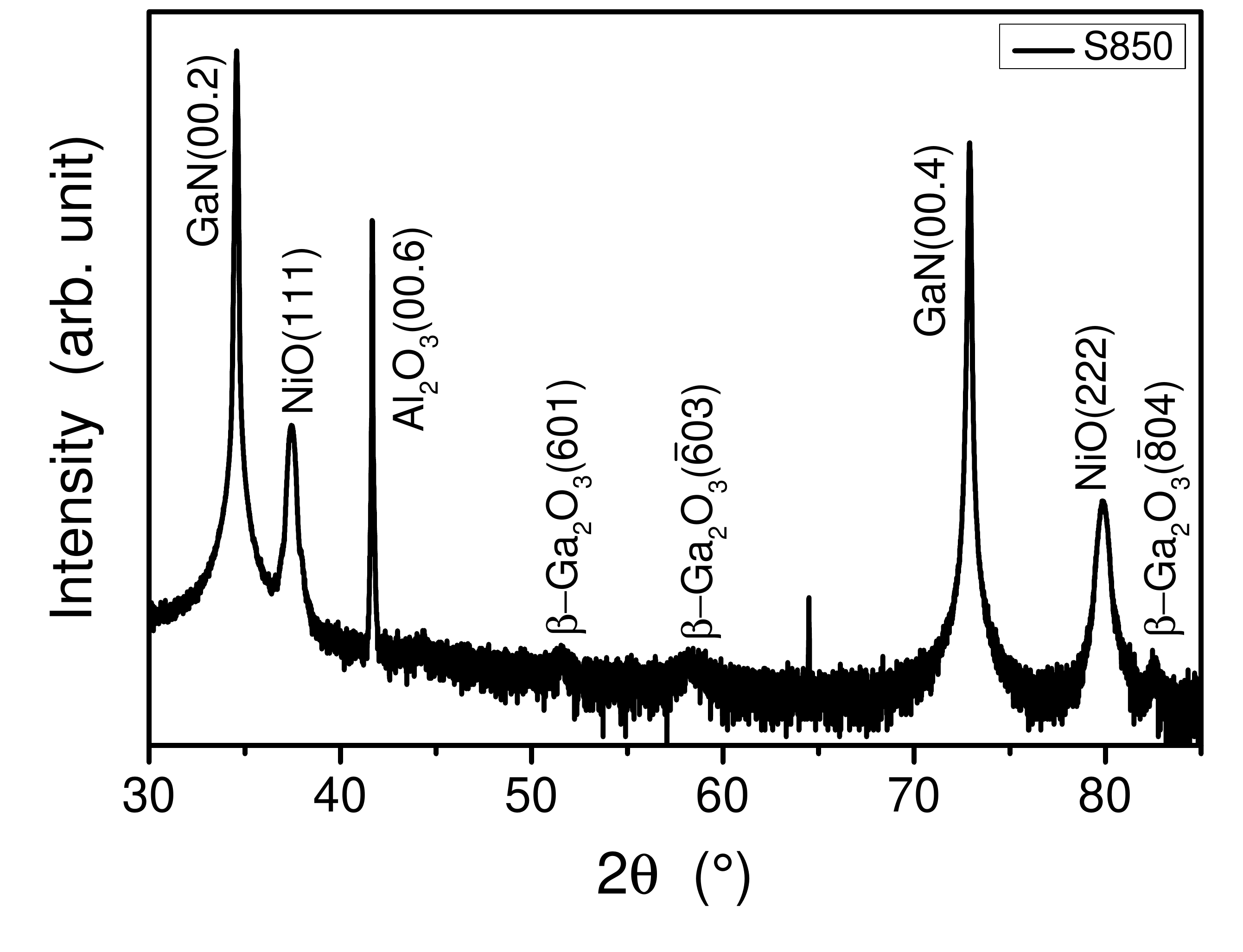}
\par\end{centering}
\caption{The XRD on-axis $2\theta-\omega$ scan of S850 showing additional
weak peaks fitting to $\mathrm{\beta-Ga_{2}O_{3}}$. The AlN buffer
peaks are missing for this charge of GaN wafers.\label{fig:950-XRD}}
\end{figure}

\begin{figure}
\begin{centering}
\includegraphics[width=8cm]{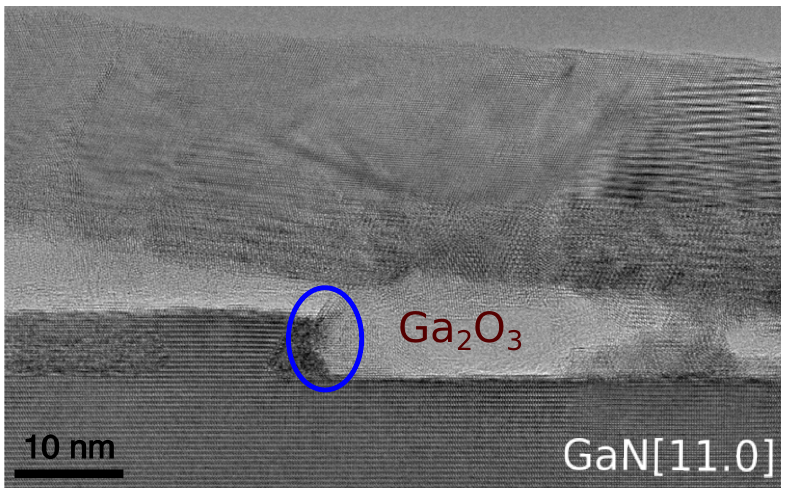}
\par\end{centering}
\caption{Cross-sectional TEM images of the S850. The bright contrast shows
the Ga$\mathrm{_{2}}$O$\mathrm{_{3}}$ interlayer between substrate
and NiO is visible. In addition, the different thicknesses of the
interlayer (blue ellipse) shows the decomposition of the substrate.\label{fig:TEM-images-of-950}}
\end{figure}

During oxide growth, unintentional oxidation of the substrate can
lead to an additional interfacial oxide layer at the substrate surface.
For example, Tsai et al. have suggested the formation of an insulating
Ga$\mathrm{_{2}}$O$\mathrm{_{3}}$ layer during PAMBE of Sb-doped
SnO$_{2}$ at increased growth temperature as cause for the drastically
increased contact resistance to the underlying GaN-based LED structure.\citep{Tsai2015}
As explained in the experimental section of our study, a layer of
Ni was grown in the first minute without oxygen to avoid this problem.
The XRD measurements of S100--S700 (cf. Fig.~\ref{fig:2Theta Combination})
did not show any signature of Ga$\mathrm{_{2}}$O$\mathrm{_{3}}$
nor did the TEM images show an interfacial oxide layer. At the highest
investigated growth temperature, 850~$^{\circ}$C for S850, however, the $\mathrm{2\theta-\omega}$
scan (see Fig.~\ref{fig:950-XRD}) shows additional features around
58.38$^{\circ}$ and 82.52$^{\circ}$. These can be assigned to the ($\mathrm{\bar{6}03}$)
and ($\mathrm{\bar{8}04}$) reflections of the ($\mathrm{\overline{2}}$01)-oriented
\textgreek{b}-$\mathrm{Ga_{2}O_{3}}$ -- the same orientation that
has been observed for intentionally grown \textgreek{b}-$\mathrm{Ga_{2}O_{3}}$
films on GaN(00.1) by Nakagomi et al.\citep{Nakagomi2015b} The ($\mathrm{\bar{2}01}$)
and ($\mathrm{\bar{4}02}$) peaks are not visible in our scan. However,
the ($\bar{4}02$) could be hidden under the NiO(111) shoulder. (An
additional reflection around 51.67$^{\circ}$ in S850 could be related to (601)
oriented $\mathrm{\beta-}$$\mathrm{Ga_{2}O_{3}}$.) TEM images confirm
the formation of a $\mathrm{\approx10}$\,nm-thick, inhomogeneous,
interfacial $\mathrm{Ga_{2}O_{3}}$ layer in S850 as shown in Fig.~\ref{fig:TEM-images-of-950}.
Furthermore, the TEM image shows a local decomposition of the GaN
substrate, responsible for a rough GaN and NiO surface.

A reference growth of NiO at 500\,$^{\circ}$C \emph{without} initial Ni deposition
(S500{*}), did not show interfacial $\mathrm{\mathrm{Ga_{2}O_{3}}}$
in the HRTEM image (see Fig.~S2), either. These results suggest the
formation of an interfacial Ga$\mathrm{_{2}}$O$\mathrm{_{3}}$\nobreakdash-layer
to be caused by the decomposition of GaN as a result of elevated growth
temperature (and subsequent oxidation of the free Ga) rather than
the pure presence of oxygen. Fern\'{a}ndez\nobreakdash-Garrido et al.
have shown a GaN decomposition in vacuum to start at surface temperatures
of about 700~$^{\circ}$C,\citep{GaN_desorption} which would mean lower growth
temperatures of the NiO layer can prevent the formation of interfacial
Ga$\mathrm{_{2}}$O$\mathrm{_{3}}$. Regarding the layer quality,
Raman results (cf. Fig.~\ref{fig:Raman}), as well as the FWHM (cf.
$\mathrm{\triangle\omega}$ in Tab.~\ref{tab:size+d-of-AFM+XRD})
show similar values compared to S500. The values even indicate a slight
enhancement of the quality without a Ni preflow. Especially the FWHM,
shows the lowest value of 0.152 comparable to the one of S100. On
the other hand, an increase of the roughness and islands size is observed.
In summary, besides increased layer roughness neither reduction of
the quality nor formation of a Ga$\mathrm{_{2}}$O$\mathrm{_{3}}$
interlayer was found for S500{*}. Since a Ni preflow is not needed
to avoid the oxidation of the GaN substrate and can even lead to surface
traps in it,\citep{Ni_GaN_Trap} we do not recommend a Ni preflow
as it was used in this study for most samples.

\noindent 

\section{Summary and conclusions}

NiO growth by PA\nobreakdash-MBE on GaN(00.1) was studied at growth
temperatures ranging from 100\,$^{\circ}$C to 850\,$^{\circ}$C. XRD measurements of
the films showed the formation of epitaxial NiO(111) for all growth
temperatures. As expected for the growth of layers with threefold
rotational symmetry on a sixfold rotationally symmetric substrate,
two types of columnar, 10--25\,nm-wide rotational domains, rotated
by 60$^{\circ}$ with respect towards each other were identified by XRD, RHEED,
TEM, and EBSD. The resulting out-of-plane and in-plane epitaxial relations
of cubic NiO on wurtzite GaN are: NiO(111)$\mathrm{\:||\:}$GaN(00.1)
and $\mathrm{NiO(1\overline{1}0)\:||\:GaN(11.0)}$ (for domain 1)
or $\mathrm{NiO(10\overline{1})\:||\:GaN(11.0)}$ (for domain 2).
In contrast, on the isostructural cubic MgO(100) substrate, NiO(100)
grows without rotational domains, demonstrating the importance of
symmetry for epitaxial growth. The layers on GaN are nearly relaxed
with an in\nobreakdash-plane tensile strain below 1\,\%. TEM confirms
layer relaxation at the interface by the formation of closely-spaced
a dislocation network. Even though the mismatch between NiO and GaN
is 8.1\,\%, a growth by domain-matching-epitaxy can neither be proven
nor excluded as the relative small domain sizes hinders an analysis
of the threading dislocation densities. A clear formation of interfacial
Ga$\mathrm{_{2}}$O$\mathrm{_{3}}$ was found at the highest growth
temperature of 850\,$^{\circ}$C whereas no Ga$\mathrm{_{2}}$O$\mathrm{_{3}}$
was detected at growth temperatures of 700\,$^{\circ}$C and below, suggesting
thermal decomposition of the GaN substrate as the cause of interfacial
Ga$\mathrm{_{2}}$O$\mathrm{_{3}}$ formation rather than the sole
exposure of the GaN to oxygen plasma. AFM measurements showed a 3D
dominated surface with islands increasing in size with increasing
growth temperature. The film grown at 850\,$^{\circ}$C was significantly rougher
than all other films as a result of the inhomogeneous interfacial
oxidation of the GaN.

For quality evaluation the area ratio between the forbidden one\nobreakdash-phonon
Raman peak and the allowed two\nobreakdash-phonon Raman peak was
used, indicating the lowest concentration of parity-breaking defects
for S700. Compared to our results for NiO grown on isostructural cubic
(and almost lattice matched) MgO(100), the quality is drastically
reduced, likely due to the domains and related domain boundaries.
A frequently used indicator of dislocations is the XRD $\omega$-rocking
curve of the film peak, which is typically wider for a larger concentration
of dislocations due to the increased tilt mosaic they cause.\citep{Itoh1988}
All investigated films have very similar rocking curve widths with
the narrowest ones exhibited by S100 and S500{*}.\\
Fern\'{a}ndez\nobreakdash-Garrido et al.\citep{GaN_desorption} identified
700\,$^{\circ}$C as the critical temperature for GaN desorption, giving it
as an upper limit for high quality layers on this substrate. This
situation is similar, to the growth on MgO, where the onset of magnesium
diffusion from the substrate into the NiO film defined an upper limit
of the growth temperature.\citep{MgO_NiO} A lower limit was not found
on both substrates, showing epitaxial NiO layers even down to 20~$^{\circ}$C
for the investigations on MgO. In addition, on GaN the lowest roughness
and XRD $\omega$-rocking curve widths was measured for 100\,$^{\circ}$C,
which could be beneficial for devices at the expense of lower grain
sizes and higher density of parity\nobreakdash-breaking defects.

\subsection*{SUPPLEMENTARY MATERIAL}

See supplementary material for a detailed discussion about the analysis
of the EBSD pattern and for TEM images from S100 and S500{*}.
\begin{acknowledgments}
\noindent We would like to thank H.-P. Sch\"onherr for MBE support,
O. Brandt and V. Kaganer for helpful discussions. This work was performed
in the framework of GraFOx, a Leibniz ScienceCampus partially funded
by the Leibniz Association. P.F, J.F., and M.B. gratefully acknowledge
financial support by the Leibniz Association. Providing beamtime at
the PHARAO endstation of BESSYII (Helmholtz-Zentrum Berlin) is appreciated.
\end{acknowledgments}

\bibliographystyle{unsrtnat}
\bibliography{GaN+NiO_paper}

\clearpage
\widetext
\begin{center}
\textbf{\large Supplemental Materials: Plasma-assisted molecular beam epitaxy of NiO on GaN(00.1)}
\end{center}
\setcounter{equation}{0}
\setcounter{figure}{0}
\setcounter{table}{0}
\setcounter{page}{1}
\makeatletter
\renewcommand{\theequation}{S\arabic{equation}}
\renewcommand{\thefigure}{S\arabic{figure}}

\subsection*{S1 Electron Backscatter Diffraction}

\noindent In EBSD measurements, the crystal orientation can be mapped
over the sample surface. Fig.~S\ref{fig:EBSD}(a) shows an in-plane
orientation map obtained by an automatic indexing of the Kikuchi patterns
for the NiO film grown at 300~$^\circ$C with a step size of 20~nm.
However, the Kikuchi patterns are recorded with a reduced resolution
to speed up the acquisition of such a map, which prevents a drift
of the sample during the measurement. Therefore, patterns with common
bands, as it is the case for twin orientations, might be harder to
distinguish. The map contains grains with diameters of up to about
500~nm. White pixels were misindexed with different orientations
due to a low pattern quality. Overall, the blue orientation shows
a higher occurrence and also exhibits larger grain sizes than the
red orientation. However, for the small grain sizes evidenced by TEM,
the Kikuchi patterns can consist of a superposition of both twin orientations,
where the automatic indexing procedure will decide for one of the
two contained orientations. Indeed, the average confidence index for
the map is only 0.09, which confirms that the algorithms decision
for on of the twin orientations is quite ambiguous. The algorithm
decides between several possible orientations by a voting mechanism
and such a low confidence index ($CI$) indicates that two possible
orientations received a similar number of votes ($V_1$ and $V_2$),
as $CI=(V_1-V_2)/V_{max}$. In consequence, EBSD is able to confirm
the presence of the two in-plane orientations, but the average grain
size is below the limit to reliably map the distribution of the twin
orientations of the NiO film by this technique.

\noindent 
\begin{figure}[b!]
\noindent \begin{centering}
\includegraphics[width=8cm]{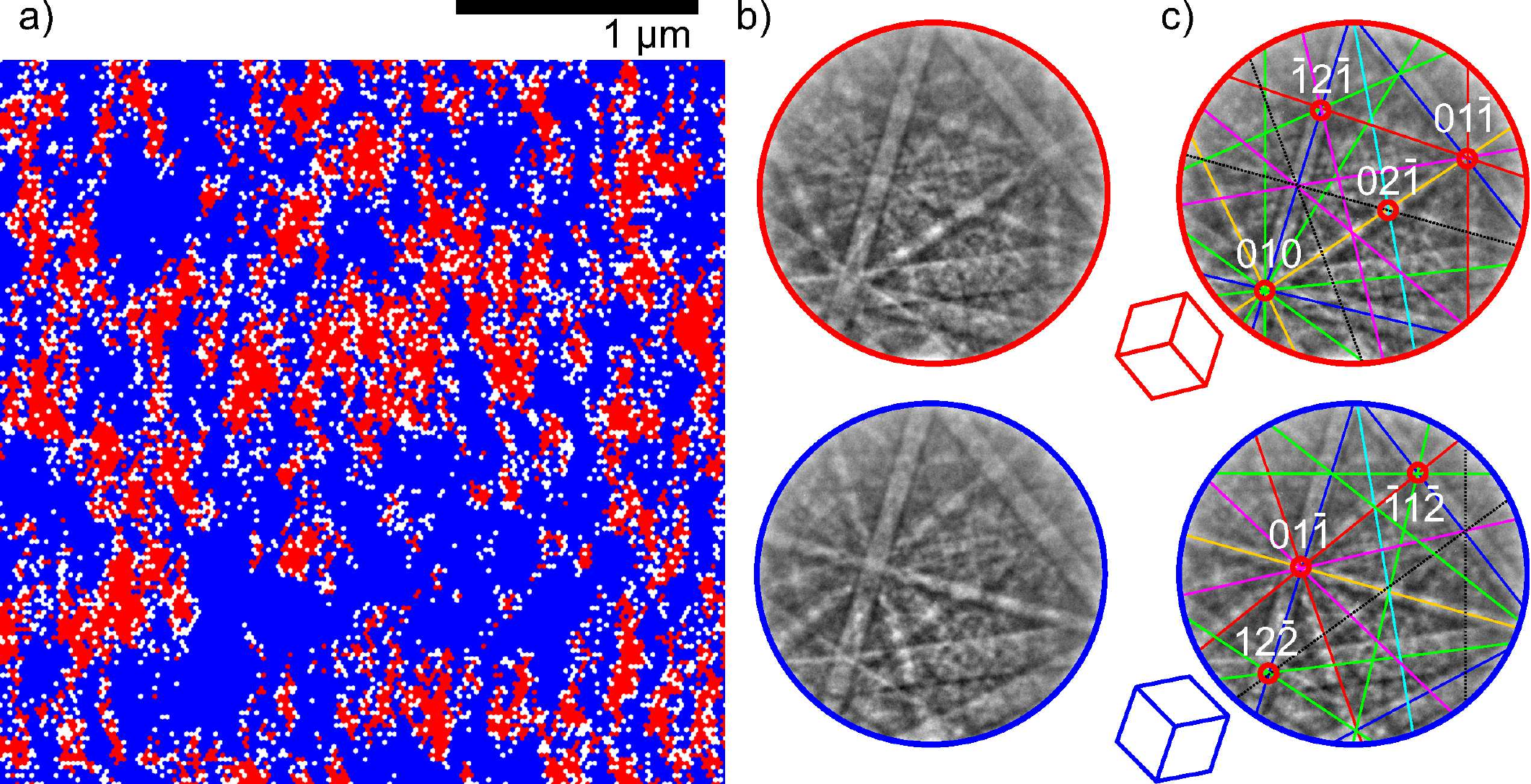}
\par\end{centering}
\caption{(a) Map of the in-plane orientation obtained from EBSD measurements
with a step size of 20~nm. The red and blue domains correspond to
the two twin orientations rotated by 60$^{\circ}$. The white pixels were misindexed
due to a low pattern quality. (b) Examples of the respective Kikuchi
patterns for the two domains and (c) indexed Kikuchi patterns as shown
in the main manuscript.\label{fig:EBSD}}
\end{figure}

\subsection*{S2 Transmission Electron Microscopy}

\begin{figure}
\begin{centering}
\includegraphics[width=8cm]{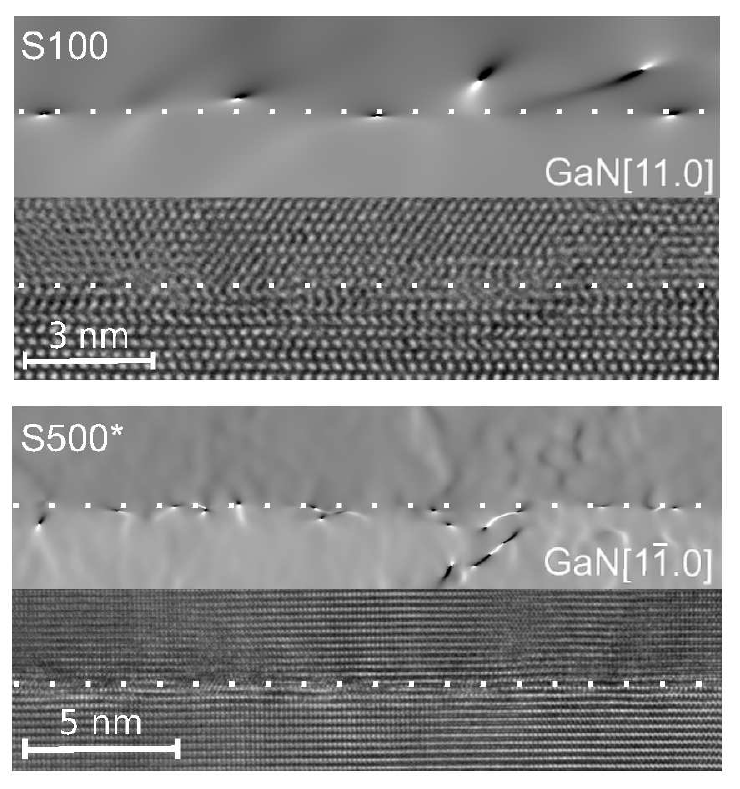}
\par\end{centering}
\caption{The strain maps and corresponding HRTEM images of the samples S100
(GaN{[}11.0{]} projection) and S500{*} (GaN{[}1$\mathrm{\overline{1}}$.0{]}
projection). The white dotted lines are a guide for the eye and indicate
the interface between substrate and NiO layer.\label{fig:TEM-images}}
\end{figure}

\noindent The HRTEM images of S100 and S500{*} (grown at 500 $^{\circ}$C without
an initial Ni deposition) show no indication of a $\mathrm{Ga_{2}O_{3}}$
interlayer. The strain maps for both samples are also shown in Fig.~S\ref{fig:TEM-images},
showing a distinct dislocation network. The lower periodicity of the
dislocation network in S100 could indicate lattice match epitaxy (LME).
A higher periodicity, on the other hand, as found in Fig.~8 for S700
underlines domain matching epitaxy (DME). However, this effect can
also be attributed to distortion relaxations which can occure at the
surface and vary locally. Accordingly, as already suggested by the
different strains, a change from LME to DME with higher temperature
is possible, but no clear conclusions can be drawn from our results.

\end{document}